\documentclass[seceq]{ptptex}

\title{
An Interpretation of ``Superluminal Neutrinos" Compatible with Relativity 
in the Framework of Standard Model
\footnote{Essential contents of the present paper have appeared already
in a paper written in Japanese\cite{rf:1}.}
}

\author{
Noboru \textsc{NAKANISHI}
\footnote{Professor Emeritus of Kyoto University. E-mail: nbr-nak@trio.plala.or.jp}}
\inst{
\textit{12-20 Asahigaoka-cho, Hirakata 573-0026, Japan}
}

\abst{
According to the measurement of muon-neutrino experiment done
by the OPERA collaboration, the speed of high-energy neutrinos
exceeds that of light in vacuum by 25ppm. Assuming that this result is
correct, a possible resolution of the dilemma between it and the validity of 
relativity is proposed
without changing the framework of the Standard Model of elementary
particles. The essential idea is based on a possible resolution,
proposed previously, of the color 
confinement problem of quantum chromodynamics.
}

\begin{document}

\maketitle

\section{Introduction}

Very recently, the OPERA collaboration\cite{rf:2}
has made a very sensational claim that
according to their precision measurement, high-energy muon-neutrinos  
traveled from CERN to the underground Gran Sasso Laboratory
(apart from CERN by 730km)
\textit{faster} than the speed of light
measured in vacuum. Let $v$ and $c_0$ be the speed of their neutrinos and that
of light measured in vacuum, respectively; their result reads
\begin{equation}
(v-c_0)/c_0=[2.48\pm0.28(\mathrm{stat.})\pm0.30(\mathrm{sys.})]\times 10^{-5}.
\end{equation}

Although the classical theory of relativity does not forbid the existence 
of the faster-than-light particles, called ``tachyons", which have a purely
imaginary mass, it is known that the tachyon field cannot be consistently 
quantized in the framework of the positive-metric Fock space\cite{rf:3}. Furthermore,
from the experimental setting, it is impossible to interpret the OPERA
neutrinos as tachyons. Hence, if their neutrinos have traveled truly faster than
the speed of light in vacuum, the OPERA experimental result is incompatible
with relativity. However, since the validity of relativity is supported
by a vast amount of experimental evidences and by the brilliant successes of theories
formulated on the basis of relativity, it is totally inacceptable that such a single 
experimental result as above rules out relativity. 
The only way-out is to assume that the fundamental constant $c$ of relativity,
``the speed of light in the true vacuum", is  different from $c_0$. Then,
everything is all right if $c_0<v<c$.

The problem is then why $c_0$ is smaller than $c$. As is well known, light 
travels in a medium slower than in vacuum. Therefore, if there exists an
undetectable medium on Earth, then the measured speed, $c_0$, of light 
in such a medium is smaller than the speed, $c$, of light in the true vacuum.
A candidate of such a medium is dark matter. However, if dark matter consists
of neutral elementary particles, such as neutralinos assumed in SUSY, then
it cannot play the role of a medium, because a medium must consist of 
charged particles at the microscopic level 
as the ordinary matter does. Previously, the present
author proposed to explain dark matter by the neutron-like cloud of dispersed
quarks\cite{rf:4}. Since quarks are charged particles, this dark-matter candidate can
be regarded as a medium. In the present paper, however, we concentrate our attention
only to the problem of finding the resolution of the OPERA neutrino problem,
forgetting the problem of the total amount of dark matter. 

In Section 2, we review the theory of 
unobservable quarks. In Section 3, we discuss the compatibility of our
proposal with the observation of the supernova SN1987A. In Section 4,
some comments are made on related works.

\section{Theory of unobservable quarks}

According to the Standard Model, hadrons consist of quarks (and antiquarks),
but no quarks are observable. Quarks have a color charge, but experimentally
it is known that only colorless states are observable. The dynamics of quarks
is successfully described by quantum chromodynamics (QCD), but the problem
of color confinement is usually regarded as unsolved yet. The conventional
approach to this problem is only to discuss the confinement of the
quark-antiquark system dynamically. The color confinement is, however,
quite a universal phenomenon relevant to any complicated hadron-like system.
It is quite unnatural to expect that such a clear-cut qualitative property
as color confinement
can be established only by means of complicated calculations of dynamics.
In 1984, therefore, the present author and Ojima\cite{rf:5}
proposed a very simple way of color
confinement in the framework of QCD. We review this theory very briefly in
the following.  

Of course, QCD is a non-abelian gauge theory. It is well known that the 
introduction of an indefinite-metric Hilbert space is indispensable in the 
manifestly covariant quantization of gauge theory. In order to secure the
probabilistic interpretability, one must set up a subsidiary condition to
define the physical subspace. The manifestly covariant formalism of a
non-abelian gauge field is most satisfactorily formulated by using the
BRS invariance. Then, according to the Noether theorem, the BRS charge $Q_B$
exists. The physical subspace $\{|\mathrm{phys} \rangle \}$ is defined by
the Kugo-Ojima subsidiary condition
\begin{equation}
Q_B |\mathrm{phys} \rangle = 0.
\end{equation}

Now, the proposal made in the previous paper\cite{rf:5} is as follows.
QCD is an $SU(3)$-invariant gauge theory; correspondingly, there exists
the color charge $Q^a$, where $a$ denotes the color index. In addition to
(2.1), we introduce an extra subsidiary condition
\begin{equation}
Q^a |\mathrm{phys} \rangle = 0.
\end{equation}
Because $[Q^a, Q_B]=0$, (2.2) is compatible with (2.1). Under this setting-up,
all colored states are unphysical; hence they should be unobservable.

Although color confinement has been \textit{globally}
realized by the above setting-up, this does not necessarily imply that
it is so \textit{locally}. There is the problem of ``behind-the-moon". 
Suppose that we have wished to make all electrically charged states unphysical
by introducing a subsidiary condition $Q|\mathrm{phys} \rangle = 0$,
where $Q$ is the electric charge. But this necessarily becomes unsuccessful: 
For example, consider a state consisting an electron at hand and a positron
behind the moon. Then we can practically have an electron only. 
Mathmatically, our physical state is a 
\textit{direct product} of the electron state whose
wave packet is located here and the positron state whose wave packet is located
behind the moon. However, this is the speciality of the abelian symmetry.
Since $SU(3)$ is a non-abelian symmetry, there is quantum entanglement in
a many-particle state; that is, any colorless state cannot be a direct product
of several colored states consisting of unoverlapping wave packets.
 Precisely speaking, we can prove the following theorem:
``Let $\varphi_1$ and $\varphi_2$ be the local operators whose supports
are restricted to spacetime domains $\varOmega_1$ and $\varOmega_2$,
respectively, which are mutually spacelikely separated. Furthermore,
let $|0\rangle$ be the vacuum state and $Q^a$'s be the charge operators
of a semisimple group. If $Q^a \varphi_1 \varphi_2
|0\rangle =0$, then we have $Q^a \varphi_1 |0\rangle =0$ and 
$Q^a \varphi_2 |0\rangle =0$."
Therefore, any colored particle is unobservable under the subsidiary
condition (2.2). Detailed 
mathematical considerations based on the gauge-theoretical
extension of the axiomatic quantum field theory are presented in a 
previous paper\cite{rf:6}. 

Thus, in high-energy reactions involving hadrons, isolated
quarks (and anti-quarks) may be created, but \textit{they cannot be observed
as particles}. In this sense, color confinement is achieved.
Since the physical subspace no longer has the Fock structure
with respect to asymptotic fields, the cross
section of an inclusive reaction can be truly larger than the total sum of 
the corresponding exclusive reactions, because there may exist the
reactions emitting unobservable quarks\footnote{But the converse is not
possible because we cannot prepare the physical state consisting of
dispersed quarks}. 

At the beginning of the universe,
a large amount of such unobservable quarks might be created. They would 
constitute, so to speak, ``cloud" of dispersed quarks. It should be electrically 
neutral in the large scale, though each quark is charged.
It has gravitational interaction in the same way as the ordinary matter, 
and hence will behave like dilute gas at the macroscopic level. 
On the other hand, it has electromagnetic interaction at the microscopic level,
but its existence cannot be detected by the usual spectroscopy.

\section{Constraint from the supernova SN1987A}

Now, our hypothesis is that we are surrounded by the cloud of dispersed quarks,
which play the role of an undetectable medium of light. What we measure as the speed
of light in vacuum is nothing but the speed $c_0$ of light in this medium. 
In this section, we examine whether or not our hypothesis is compatible with the
observation of the supernova SN1987A.

As is well known, in 1987, neutrinos coming from the supernova,
named SN1987A, in Large Magellanic Cloud were observed in KAMIOKANDE\cite{rf:7}
and two other places.
In spite of the distance of about $1.68\times 10^5$ light-years,
SN1987A was optically observed
only \textit{3 hours later}. From this fact, it is seen that the difference of 
the speed of neutrinos and that of light is of order $10^{-9}c$ in sharp contrast
with the OPERA data. Although the energies of the SN1987A neutrinos are 
around $E$=10MeV,
while the mean energy of the OPERA neutrinos is 17GeV, both energies are 
ultra-relativistic for neutrinos, and hence it is unnatural to suppose that
there exists an essential difference between them. Rather, it is natural to 
expect that 
the essential difference comes from the experimental situations; the SN1987A
observation is cosmological, while the OPERA experiment was done on Earth. 

If we assume that the SN1987A neutrinos with mass $m$ traveled in the 
cloud of dipersed quarks in the whole distance, from (1.1) we have
$m=E\sqrt{1-(c_0/c)^2}\simeq85\mathrm{keV}$, which is too much larger than an
upper bound of the neutrino mass. Hence, they must travel in the true vacuum in  alomst all portion of the distance. That is, we should suppose that the
clouds of dispersed quarks exist only in the neighborhoods of the 
supernova and the Solar system (on the neutrino path).
Since 3 hours=$3.4\times10^{-4}$years, an upper bound of the length of its
existence region is $3.4\times10^{-4}c/2.5\times10^{-5} \simeq 13.7$ light-years.
This value is sufficiently larger than the radius of the Solar system.
Thus, if we assume that the cloud of dispersed quarks exists only in the
regions of high matter density, our hypothesis is consistent with the observation
of SN1987A.

\section{Discussion}

We have proposed a hypothesis, on the basis of the Standard Model with the color
confinement condition (2.2), which gives a possible interpretation of
the OPERA experimental result without bringing any contradiction with the
validity of relativity. If the cloud of dispersed quarks is assumed to
exist mainly near the regions of high matter density, our hypothsis is
consistent also with the observation of SN1987A. This assumption is natural
because the cloud of dispersed quarks has the gravitational interaction exactly
in the same way as the ordinary matter does. Furthermore, the
existence of the cloud of dispersed quark will bring no
anisotoropy of the speed of light caused by the rotation of Earth
because it behaves like air (but probably much lighter) in contrast with the aether.

We emphasize that if our interpretation of the OPERA experimental result
 turns out to be correct, it will mean that the proposed
resolution of the color-confinement problem of QCD is experimentally
supported.

Finally, we make some comments related to our hypothesis.\\

(1)\; According to the experiment done in SLAC\cite{rf:8} by using a time-of-flight 
technique, the speeds of gamma ray and of electrons in the energy range
15-20GeV can be different within only $2 \times10^{-7}$ times the speed of light. 
Since this experiment
was performed in the end 1/3 part of the linear accelerator, it is natural to
understand that not only air but also the cloud of dispersed quarks were 
removed from there, that is, the measurements were made in the \textit
{true vacuum}. Hence the speed of the gamma ray, which  was not measured 
in this experiment, is not $c_0$ but $c$.
Thus the result does not contradict our hypothesis but merely 
confirms the validity of relativity. \\

(2)\; Cohen and Glashow\cite{rf:9} theoretically criticized the OPERA experimental data.
According to them, if superluminal neutrinos were really emitted from CERN,
many electron-positron pairs should have to be produced via bremsstrahlung
before arriving at the OPERA neutrino detector, so that such high-energy
neutrinos as observed could not survive. Their reasoning is based on the
possibility that the processes $\nu \to \nu + \mathrm{something}$
are kinematically allowed if the neutrino is really superluminal. 
Their criticism is, however, no longer applicable if $c_0<v<c$, because
then we can always take the rest frame of the neutrino. A similar comment 
has been made independently also by Oda and Taira\cite{rf:10} \footnote{The author would 
like to thank Prof. I. Oda for this information.}.\\

(3)\; A possible energy dependence of photon speed was observed
in an experiment done several years ago\cite{rf:11}.
According to the claim based on the missing energy
observation in HERA Compton polarimeter data, photons with 12.7GeV energy
were moving faster than light by 5.1(1.4)mm/sec. On the other hand, no
energy dependence of the speed of light was detected in the observation, done
by Fermi Gamma-ray Space Telescope (in 2009) \cite{rf:12},
of the gamma-ray burst which had taken place at a distance of $7.3\times
10^{9}$ light-years. 
Because an extremely small difference in the photon speeds would induce a
significantly large difference between the times of arrival
when accumulated over a tremendously remote distance,
this result is a very severe denial of the energy dependence of the photon speed
in vacuum. It is interesting to note, however, that
both observations are compatible if one supposes that there exists
such a medium as the cloud of dispersed quarks in the neighborhood of Earth.\\


The author would like to thank Prof. T. Kugo for discussing the compatibility
of our hypothesis with the SN1987A data.

%


\end{document}